\documentclass{mn2e}
\usepackage{epsfig}
\usepackage{times}
\begin{document}

\title[Cyg X-1] {
Quenched millimetre emission from Cygnus X-1 in a soft X-ray state
}
\author[S.P. Tigelaar et al.]  {S.P. Tigelaar$^1$, R. P. Fender$^1$, 
R.P.J. Tilanus$^2$, E. Gallo$^1$, G.G. Pooley$^3$ \\ \\ 
$^1$Astronomical Institute
`Anton Pannekoek', University of Amsterdam, and Center for High Energy
Astrophysics, Kruislaan 403, \\ 1098 SJ, Amsterdam, The Netherlands
{\bf rpf@science.uva.nl}\\ $^2$Joint Astronomy Centre, Hilo, Hawaii, USA 
\\ $^3$Mullard Radio Astronomy
Observatory, Cavendish Laboratory, Madingley Road, Cambridge CB3 OHE\\
}

\maketitle

\begin{abstract}

We present millimetre wavelength observations of the black hole
candidate X-ray binary Cygnus X-1 which indicate a suppression, or
quenching, of the emission as the source switches to a softer X-ray
state. Combining the data with those for another black hole candidate,
XTE J1118+480, we demonstrate that the millimetre emission shows the
same coupling to X-rays as the radio emission, although with a much
stronger sensitivity to spectral shape. We therefore confirm the
association of the millimetre emission with the jets in low/hard state
black hole candidate X-ray binaries.

\end{abstract}


\section{Introduction}

In recent years a clear connection between the radio and X-ray
emission from galactic black hole and neutron star X-ray binaries has
been established (e.g. Harmon et al. 1995, 1997; Fender et al. 1999;
Mirabel \& Rodr\'\i guez 1999; Klein-Wolt et al. 2002; Corbel et al. 2003;
Gallo, Fender \& Pooley 2003; Fender 2001, 2004). This connection is
interpreted as reflecting the coupled processes of accretion and
ejection in these systems, and analogous behaviour may be observed
from supermassive black holes in active galactic nuclei on longer
timescales (Marscher et al. 2002; Merloni, Heinz \& di Matteo 2003;
Falcke, K\"ording \& Markoff 2003; Maccarone, Gallo \& Fender 2003),
highlighting the importance of such studies.

Specifically, in the `low/hard' X-ray state, characterised by an X-ray
spectrum peaking at $\sim 100$ keV, little accretion disc emission and
strong variability (e.g. McClintock \& Remillard 2004),
there seems to be always a steady radio jet (Fender 2001). This jet is
characterised by a `flat' or `inverted' ($\alpha \sim 0$ or $\alpha >
0$ respectively, where the spectral index $\alpha = \Delta \log
S_{\nu} / \Delta \log \nu$, i.e. $S_{\nu} \propto \nu^{\alpha}$) radio
spectrum, and the (GHz) radio luminosity seems to be correlated with
the X-ray luminosity as $L_{\rm radio} \propto L_{\rm X}^b$ where $b
\sim 0.7$ (Corbel et al. 2003; Gallo, Fender \& Pooley 2003).

The radio spectrum is further observed to extend smoothly through the
millimetre bands to the near infrared based upon which it has been
suggested that the emission in these bands can also be dominated by
synchrotron emission from jets in the low/hard state.  (Fender 2001,
2004; Fender et al. 2001; Corbel \& Fender 2003; Chaty et al. 2003)
When sources transit to disc-dominated `high/soft' or `intermediate'
states at higher bolometric luminosities (Homan et al. 2001;
McClintock \& Remillard 2004) the radio emission is strongly
suppressed or `quenched' (Tanabaum et al. 1972; Fender et al. 1999;
Corbel et al. 2001; Gallo, Fender \& Pooley 2003).

The millimetre properties of X-ray binaries have been well studied
only for the brightest of sources (e.g. Baars et al. 1986; Paredes et
al. 2000). These sources tend to be semi-continuously flaring and the
association of the millimetre emission with X-ray state is hard to
determine, although millimetre oscillations are clearly associated
with rapid quasi-periodic accretion state changes in GRS 1915+105
(Fender \& Pooley 2000). Fender et al. (2000) showed that the radio
spectrum of Cygnus X-1 extended smoothly with a very flat spectrum
($|\alpha| \leq 0.15$ ($3\sigma$)) into the mm band. Fender et
al. (2001) presented observations of the transient BHC XTE J1118+480
in which a rather steeply inverted radio spectrum ($\alpha \sim +0.5$)
connected smoothly to detections in the mm band. Furthermore, once
this source had faded the mm emission had also clearly declined,
strengthening the inferred connection with the jet component. This
behaviour was consistent with the correlated radio:X-ray behaviour
observed in this state. In this paper we report the first evidence for
the quenching of millimetre emission from a black hole binary, Cygnus
X-1, when it enters a softer X-ray state. 

\section{Observations and results}

The analysis in this paper uses published millimetre fluxes for Cygnus
X-1 (Fender et al. 2000) and XTE J1118+480 (Fender et al. 2001), plus
a new observation. This was performed on 2002 April 27 16:00--18:00 UT
(MJD 52391) with SCUBA (Holland et al. 1999) on the JCMT, operating at
350 GHz, and failed to detect Cygnus X-1 with a $3\sigma$ upper limit
of 7.5 mJy. Note that although the observations of Cyg X-1 range from
89 -- 350 GHz we compare them as if the spectrum was perfectly flat
($\alpha = 0$), based upon Fender et al. (2000). This means that for
MJD 50964 we have averaged the 146 and 221 GHz measurements. While the
radio--millimetre spectrum of XTE J1118+480 (Fender et al. 2001 and
see below) indicates that assuming a flat spectrum is not always
appropriate, we believe that for the study undertaken here it is a
reasonable enough approximation.

We compare our millimetre measurements with contemporaneous X-ray and
radio monitoring of Cygnus X-1.  For X-rays we have used the publicly
available {\rm Rossi}XTE all-sky-monitor (RXTE ASM; Levine et
al. 1996) data for Cygnus X-1. These data measure the flux from the
source in the interval 1.5--12 keV.  For all of the epochs except MJD
50945 we have used the average of the individual RXTE ASM dwell
measurements over an interval of $\pm 0.25$ days from the time of the
millimetre observation. Around MJD 50945 the RXTE ASM sampling is very
sparse and we have used the average count rate over five days centred
on our millimetre observation. Inspection of the radio and X-ray light
curves on longer timescales around the times of our mm observations
gives us no reason to doubt that our interpretation of the state
behaviour as presented here is correct.  For the radio monitoring we
have used 15 GHz radio flux densities obtained at the Ryle
Telescope. More information on the radio monitoring of X-ray binaries
with the Ryle Telescope may be found in Pooley \& Fender (1997); the
quenching of the radio emission in soft X-ray states is presented in
Gallo, Fender \& Pooley (2003). The measurements in all three bands
are summarised in Table 1.

\begin{table}
\caption{ Millimetre and X-ray observations of Cygnus X-1. Apart from
the most recent, details of the millimetre observations can be found
in Fender et al. (2000). The millimetre observations were obtained at
a variety of frequencies with the JCMT, the radio data were all
obtained at 15 GHz with the Ryle Telescope, and the X-ray fluxes are
in 1.5--12 keV count rates obtained with the Rossi XTE all-sky
monitor. These data, plus X-ray hardness ratios, are plotted in Fig
1.}
\begin{center}
\begin{tabular}{ccccc}
\hline
MJD & $\nu_{\rm mm}$ (GHz) & $S_{\rm mm}$ (mJy) & $S_{\rm cm}$ (mJy) &
$F_{\rm X}$ (ct/s)\\
\hline 
50644 &89 &15.9$\pm$4.9 & $10.9 \pm 2.5$ & $32 \pm 2$ \\
50945 &146&9.2$\pm$3.5 & $6.3 \pm 2.5$ & $20 \pm 3$ \\
50950 &146 &5.8$\pm$3.2 & $12.0 \pm 2.5$ & $16 \pm 1$ \\
50960 &221 &11.6$\pm$2.1 & $11.5 \pm 2.5$ & $23 \pm 2$ \\
50964 &146/221 &14.4$\pm$ 3.6 & $12.7 \pm 2.5$ & $25 \pm 2$ \\
52391 &350 &0.7$\pm$ 2.5 & $0.5 \pm 2.5$ & $69 \pm 2$ \\
\hline
\end{tabular}
\label{mmmeasurements}
\end{center}
\end{table}

In Fig 1 we plot the millimetre, radio and X-ray hardness ratio HR1 as
a function of X-ray count rate. The HR1 is also from the Rossi XTE
monitoring data and is the ratio of counts in the energy range 3--5
keV to that in the range 1.5--3 keV.  In the `low/hard' state the mm
and X-ray points are clearly correlated, although the errors are
large. Quantitatively, excluding the measurement in the softer state
(see below) the best power-law fit to the millimetre-X-ray correlation
is of the form $F_{\rm mm} = (0.2 \pm 0.2) F_{\rm X}^{1.2 \pm 0.3}$,
steeper at the $\sim 2\sigma$ level than the $L_{\rm radio} \propto
L_{\rm X}^{0.7}$ found for the low/hard state (Gallo et al. 2003).
Above 30--35 RXTE ASM ct/sec the source transits from the `low/hard'
to a softer `intermediate' X-ray state (Belloni et al. 1996; Miller et
al. 2002), clearly revealed by a drop in HR1.  The well-established
suppression (`quenching') of the radio emission in the softer X-ray
state is evident, and the same effect appears to be apparent for the
mm emission. The 'quenched' millimetre point lies $\geq 10\sigma$
below the best-fit relation for the 'low/hard' state given above.  As
a caveat we note that variability of synchrotron emission from X-ray
binaries does tend to increase in amplitude at higher frequencies, and
we may just have been observing a dip in the flux unrelated to the
X-ray state of the source. However, given that the behaviour fits the
pattern observed for the radio emission, we consider this to be
unlikely. Note furthermore that the 'quenched' millimetre measurement
is at the highest frequency (350 GHz) which, were the spectrum
inverted as in XTE J1118+480 (see below), at the time of our
observations, this only strengthens the quenching effect. We conclude
that these data reveal for the first time the suppression of mm
emission in soft X-ray states,and that this demonstrates the
association of the mm emission with the jet component.

\begin{figure}
\centerline{\epsfig{file=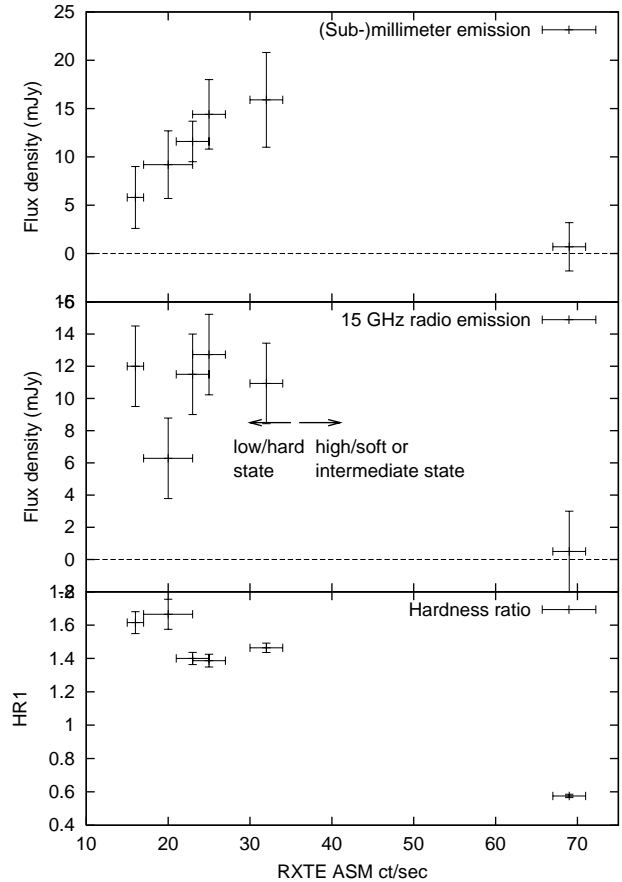, width=9cm, angle=0}}
\caption{(Sub-)millimetre (top) and radio (middle) flux densities 
and X-ray `hardness'
as a function of X-ray flux for Cyg X-1. At the highest X-ray luminosity,
the source has made a transition to a softer X-ray state, and the mm
and radio emission are `quenched'. This is the first demonstration of
suppression of the mm flux in soft X-ray states and links, as
expected, the mm emission with the jet responsible for the radio emission.}
\end{figure}

\begin{figure}
\centerline{\epsfig{file=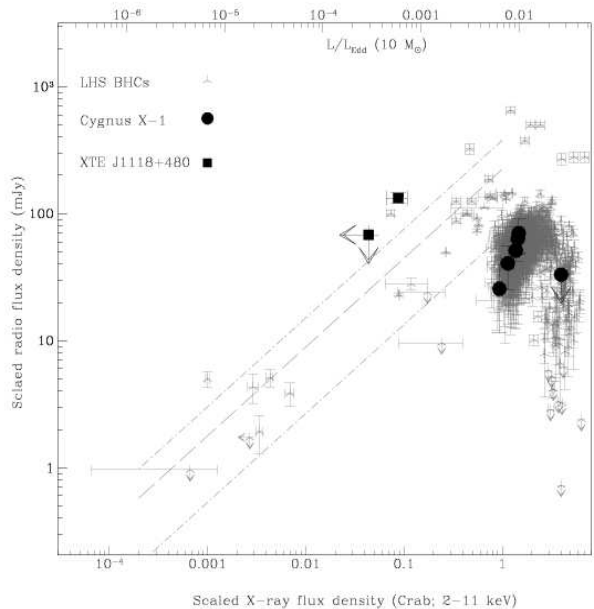, width=9cm, angle=0}}
\caption{
Millimetre flux density as a function of X-ray flux (at 1 kpc and
absorption corrected), superimposed on the radio:X-ray plane from 
Gallo, Fender \& Pooley (2003). Considering that in XTE J1118+480, the
spectrum was very steep (spectral index $\alpha \sim +0.5$), then we
expect the mm flux densities to lie about one order of magnitude above
the GHz radio data, as they do. Cyg X-1, with a flat spectrum, needs
no correction, as is also evident from the plot, which also indicates
that, just as in the radio band, the coupling in Cyg X-1 may be rather
steeper and sharper than in othe sources.
}
\end{figure}

\section{Discussion}

Cygnus X-1 was already known to have a `very flat' ($|\alpha| < 0.15$
$(3\sigma)$) spectrum from the radio (2 GHz) through mm (350 GHz)
bands while in the low/hard X-ray state (Fender et al. 2000). This was
naturally interpreted as a high-frequency extension of the
self-absorbed synchrotron spectrum from a steady jet of the kind
envisaged by Blandford \& Konigl (1979), Hjellming \& Johnston (1988),
Falcke \& Biermann (1996), and others. Direct evidence for the
extension of the synchrotron spectrum to the near-infrared has also
been found in the low/hard state black hole binary GX 339-4 (Corbel \&
Fender 2002). However, these conclusions are heavily based upon
interpretations of broad-band spectra.

In this paper we have reported the quenching of this mm emission when
Cygnus X-1 changed to a softer (high/soft or intermediate) X-ray
state, precisely as has been seen for this and other black hole X-ray
binaries in the radio band (Fender et al. 1999; Corbel et al. 2001;
Gallo, Fender \& Pooley 2003). It was already known in the case of the
black hole transient XTE J1118+480 that the mm emission dropped
significantly as the source faded (Fender et al. 2001). In Fig 2 we
present the X-ray and mm data for Cyg X-1 and XTE J1118+480,
superimposed upon the X-ray:radio diagram for all low/hard state
galactic black hole binaries from Gallo et al. (2003). The points for
Cyg X-1 lie exactly in the same `cloud' of points as the radio
emission, due to the very flat radio--mm spectrum. The points for XTE
J1118+480 are significantly above the relation, due to the strongly
inverted ($\alpha \sim +0.5$) radio--mm spectrum observed from this
source, as a result of which the flux density at 350 GHz is $\sim 10$
times that at 5 GHz. This indicates that, while the $L_{\rm radio}
\propto L_{\rm X}^{0.7}$ relation seems to hold without serious
consideration of the source spectrum, spectral effects will be
important in investigating the correlation in other bands.

The combination of the data for Cyg X-1 and that of XTE J1118+480
clearly indicates that

\begin{itemize}
\item{In the low/hard state, the mm emission correlates with the X-ray
  flux}
\item{In softer states, the mm emission is quenched}
\end{itemize}

-- this behaviour is exactly as observed for the relation between
radio and X-ray emission. 

Note that while it is clearly easier to demonstrate trends in the
disc-jet coupling by means of radio and X-ray monitoring (e.g. Gallo,
Fender \& Pooley 2003), it is of utmost importance to establish as
convincingly as possible the full spectral extent of the jet
component. Only by firmly establishing the contribution of jets to the
broadband spectrum of X-ray binaries can we accurately estimate their
contribution to the overall energetics of the system (e.g. Fender
2001). In this respect, our mm observations are important as they lie
much closer to the spectral regime in which other processes may
contribute. Bremsstrahlung emission from stellar winds (e.g. Wright \&
Barlow 1975; Panagia \& Felli 1975) and self-absorbed
(cyclo-)synchrotron emission from advection-dominated accretion flows
(e.g. Mahadevan 1997; Narayan, Mahadevan \& Quataert 1998) are
associated with spectral indices $\alpha > 0$ and may therefore be
invoked to explain the mm emission much more readily than the
radio. The coupling revealed here argues however that in black hole
X-ray binaries in the low/hard state the jet dominates the emission in
the mm band.

We therefore conclude that the mm emission arises from the same
physical component as the radio emission, most likely to be a jet-like
outflow. At such high frequencies we are probing $\sim 100$ times
closer to the base of the jet than in the radio band. The next step
will be to see if infrared emission, probing another 100-1000 times
closer, shows the same coupling to X-ray states. In this direction we
suggest that the 10$\mu$m band, where new detectors are now available,
and the thermal emission from companion stars and accretion discs
should generally be weak.

\section*{Acknowledgements}

We acknowledge with thanks the use of the quick-look X-ray data
provided by the ASM/{\it RXTE} team.  We thank the staff at MRAO for
maintenance and operation of the Ryle Telescope, which is supported by
the PPARC. The James Clerk Maxwell Telescope is operated by The Joint
Astronomy Centre on behalf of the Particle Physics and Astronomy
Research Council of the United Kingdom, the Netherlands Organisation
for Scientific Research, and the National Research Council of Canada.

\end{document}